\documentclass{ifacconf}

\usepackage{graphicx}     
\usepackage{natbib}        
\usepackage[acronym]{glossaries}
\usepackage{amssymb}
\usepackage{booktabs}
\usepackage{multirow}
\usepackage{comment}
\usepackage{xcolor}
\newcommand\numberthis{\addtocounter{equation}{1}\tag{\theequation}}
\usepackage{xurl}

\makeglossaries

\newacronym{xai}{XAI}{Explainable Artificial Intelligence}
\newacronym{mlp}{MLP}{Multi-Layer Perceptron}

\newacronym{ssl}{SSL}{Self-Supervised Learning}
\newacronym{mse}{MSE}{Mean Squared Error}
\newacronym{ssm}{SSM}{State Space Model}
\newacronym{rnn}{RNN}{Recurrent Neural Network}

\newacronym{llm}{LLM}{Large Language Model}

\newacronym{lti}{LTI}{Linear Time-Invariant}
\newacronym{afd}{AFD}{Active Fault Detection}
\newacronym{pfd}{PFD}{Passive Fault Detection}
\newacronym{glr}{GLR}{Generalized Likelihood Ratio}
\newacronym{dof}{DOF}{Degrees of Freedom}
\newacronym{mle}{MLE}{Maximum Likelihood Estimator}

\newacronym{fdi}{FDI}{Fault Detection and Isolation}

\newacronym{mm}{MM}{Minorize-Maximise}

\newacronym{rh}{RH}{Receding-Horizon}

\newacronym{sh}{SH}{Shrinking-Horizon}

\begin{document}
\begin{frontmatter}

\title{Input design for maximizing statistical detection of parametric faults in linear discrete-time systems}

\author[First]{Jeppe Roden Münster\thanksref{equal}} 
\author[Second]{Mikkel S.L. Brøndt\thanksref{equal}}
\author[Second]{Mohamad Al Ahdab} 
\author[Second]{Szymon Gres}
\author[First]{Christophe A.N. Biscio}
\author[Second]{Rafal Wisniewski}

\thanks[equal]{J.R. Münster and M.S.L. Brøndt are co-first authors.}

\address[First]{Aalborg University, Department of Mathematical Sciences, 
  Denmark (e-mails: \{jrmu, christophe\}@math.aau.dk).}
\address[Second]{Aalborg University, Department of Electronic Systems, 
  Denmark (e-mails: \{mslb, maah, sgres, raf\}@math.aau.dk)}

\begin{abstract}              
\gls{afd} concerns designing controlled inputs that in real time enhance detection and isolation of faults in dynamical systems. The \gls{afd} is typically formulated as an optimization problem to maximize a statistical distance metric between a nominal and a faulty system, subject to constraints. While numerous methods have been developed in the area of \gls{afd} to account for uncertainty in data and model parameters, many use predefined fault models, or heuristic distance metrics that have no explicit relation with system faults. In this work we propose a closed-loop approach that combines \gls{glr} test with a recursive filter that cancels out the dynamics of the monitored fault effects. The \gls{glr} test is used in an optimization scheme to design inputs that maximize the detectability of parametric faults. The proposed approach is validated on a simulation of a simple mechanical system, showcasing its effectiveness against the passive fault diagnosis methods.
\end{abstract}


\end{frontmatter}

\section{Introduction}
Fault detection and isolation are essential components in all engineering systems with numerous applications to mechanical, aerospace, automotive and wind energy systems \cite[]{6135786, 6179529, 6428675}, to name a few. To accommodate the increasing requirements for safety and performance of modern dynamical systems, there is a constant need for developing faster and more reliable methods to detect and isolate faults. A common taxonomy of fault-detection approaches distinguishes between data-driven and model-based methods. 

Among the model-based approaches, statistical fault detection methods \cite[]{book} have proven successful by inferring parametric changes of model-based residuals with robust statistical hypothesis tests. Traditionally, the statistical fault detection methods have been developed in a passive framework \cite[]{zhang2014, gao2015survey,dohler2017}, where faults are diagnosed using the observations collected under the excitation present during routine operation of the system. 

For a finite observation interval, however, the detection of small faults can be obscured by model uncertainty and data variability. Consequently, low sensitivity to faults under the available excitation can reduce the effectiveness of fault detection tests and increase the false negative test rates.

In \gls{afd} these limitations are addressed by designing inputs to improve the sensitivity of the measured outputs to faults \citep{niemann2006setup}. For linear systems with bounded uncertainties, inputs can be designed to separate a set of predefined fault models over a finite horizon, while minimizing input energy, or another control objective \cite[]{ESNAASHARI20111887, SCOTT20141580}. 

Set-valued observers can also be used to update the control inputs as new measurements become available \citep{raimondo2016closedloop} and design inputs to promote separation of the healthy and faulty output estimation sets \citep{xu2021observer}.  

For stochastic systems, auxiliary inputs have been designed by minimizing an upper bound on the probability of model misclassification \citep{blackmore2006finite} or by increasing statistical distances between predicted output distributions \citep{kim2013}. %
The aforementioned methods, however, usually assume a predefined actuator fault set, do not consider faults in the output equation, and oftentimes, assume a full knowledge of the system states, all of which are not always available in practice.

In this paper, we consider designing an input to actively detect and isolate small parametric changes in stochastic \gls{lti} systems. To the best of the authors knowledge, the proposed approach depends on weaker assumptions than those considered in the literature, particularly regarding prior knowledge of possible fault scenarios, the presence of faults in the outputs, and the use of measured outputs rather than the system states. More specifically, the faults are modelled as additive terms in system dynamics and are detected using a \gls{glr} test combined with a recursive Kalman filter that cancels out the dynamics of the faults effects, which is inspired by \cite{zhang2014}. For a fault in a specified system parameter, the input is designed to maximize the noncentrality parameter of the \gls{glr} statistic subject to a bound on its norm at each sample. When the fault direction is unknown, the input is designed to maximize the minimum expected noncentrality over a finite set of all parameter changes. Both offline and \gls{rh} solutions are considered and compared with passive detection in a simulation of a simple mechanical system.

\section{Problem Statement} \label{sec:problem}

Consider a nominal fault-free \gls{lti} stochastic discrete-time system of the form
\begin{equation}\label{eq:nom_system}
\begin{aligned}
    x_{k+1} &= A(\theta_0) x_k + B(\theta_0)u_k + \overline{B}(\theta_0)z_k + w_k,\\
    y_k &= C(\theta_0) x_k + D(\theta_0) u_k + \overline{D}(\theta_0)z_k+ v_k,
\end{aligned}
\end{equation}
where $x_k\in\mathbb{R}^n$ is the state vector, $u_k\in\mathbb{R}^l$ is a designed  input, $z_k \in \mathbb{R}^{l}$ is a measured deterministic input, $y_k\in\mathbb{R}^r$ is the measured output, and $A(\theta_0)$, $B(\theta_0)$, $\overline{B}(\theta_0)$, $C(\theta_0)$, $D(\theta_0)$, and $\overline{D}(\theta_0)$ are constant matrices of appropriate dimensions evaluated at the nominal parameter vector $\theta_0$. Additionally, we assume that these matrices are differentiable with respect to $\theta$ at $\theta_0$. For simplicity of notation, the dependence on $\theta_0$ is omitted hereafter. While $\overline{B}$ and $\overline{D}$ can differ from $B$ and $D$, they are assumed to be equal in this work. The sequences $w_k$ and $v_k$ are jointly Gaussian, zero-mean white noise processes with covariance matrices $Q$ and $R$, respectively, and cross-covariance $S$. The initial state condition $x_0$ is assumed to be Gaussian with mean $\hat{x}_0$ and covariance $P_0$. The matrix pair $(A,C)$ is assumed to be observable, and the pair $(A,B)$ controllable. 

In this work, similar to \cite{dohler2017,GRES2023}, we consider a small change $\theta_{\text{err}}=\Delta\theta+\theta_0$ in the nominal parameters such that the model in \eqref{eq:nom_system} can be approximated by a first-order Taylor expansion around $\theta_0$ to obtain
\begin{equation}\label{eq:faulty_system}
\begin{aligned}
    x_{k+1} &= A x_{k} + B \bar{u}_k
    + \Xi_k \Delta\theta + w_k,\\
    y_{k} &= C x_{k} + D \bar{u}_k
    + \Phi_k \Delta\theta + v_k,
\end{aligned}
\end{equation}
where $\Xi_k\in\mathbb{R}^{n\times p} $ and $\Phi_k\in\mathbb{R}^{r\times p}$ are the fault sensitivity matrices associated with the state and output equations, respectively, and $\bar{u}_k=u_k+z_k$. In particular, $\Xi_k=\Psi_k+\Upsilon_k$, where $\Psi_k$ and $\Upsilon_k$ are the fault signatures associated with parametric changes in $A$ and $B$, respectively.

Note that \eqref{eq:faulty_system} can be interpreted as an exact additive fault model, where $\Delta\theta$ represents the fault amplitudes and $\Xi_k$ and $\Phi_k$ are possibly time-varying fault signature matrices.

The goal of this work is to design the input $u_k$ such that the probability of detection of a parametric fault $\Delta\theta$ is maximized. Later on it is shown that this constitutes maximizing the probability of both fault detection and isolation of specific change directions. The design inputs are obtained by heuristic maximization of a noncentrality parameter of the \gls{glr} test to detect changes in $\Delta\theta$.

\section{Statistical hypothesis testing for FDI}\label{sec:StateEst}
In this section, the statistical hypothesis testing framework to detect and isolate additive faults is recalled after \cite{zhang2014}.

\subsection{Fault effect on innovation}\label{sec:KalmanFilter} 
Let $\widehat{x}_{k+1}$ be the one-step ahead Kalman prediction of the state $x_{k+1}$ for the nominal system in \eqref{eq:nom_system}
\begin{equation}\label{eq:pred_next_state}
    \widehat{x}_{k+1|k} = A \widehat{x}_{k|k-1} + B \bar{u}_k + K_k \epsilon_k,
\end{equation}
where $K_k= G_{k|k-1} \Sigma_{k|k-1}^{-1}$, $G_{k|k-1} = A P_{k|k-1} C^\top + S$, $\Sigma_{k|k-1}=C P_{k|k-1} C^\top + R$
with innovation $\epsilon_k$
\begin{align*}
    \epsilon_k &= y_k - \hat{y}_k = C \Tilde{x}_{k|k-1} + v_k,  \numberthis \label{eq:nominal_innovation}
\end{align*}
and state error $\tilde{x}_k$
\begin{align}\label{eq:difference_state}
    \Tilde{x}_{k+1|k} &= (A - K_k C) \Tilde{x}_{k|k-1}+ w_k - K_k v_k .
\end{align}

To illustrate the effect of a fault on the predicted state and the innovation, the Kalman predictor designed for the nominal system \eqref{eq:pred_next_state} is applied to the faulty system in \eqref{eq:faulty_system}.
The state error for the faulty system \eqref{eq:faulty_system} can be split into a noise-driven part $\zeta_k$ and a fault-driven part $\Gamma_k\Delta\theta$, such that
\begin{align}
    \tilde{x}_{k} = \zeta_k + \Gamma_k\Delta\theta,
\end{align}
where $\Gamma_k \in \mathbb{R}^{n\times p}$ is recursively defined by
\begin{equation}\label{eq:Gamma_next}
    \Gamma_{k+1} = (A - K_k C)\Gamma_k + \Xi_k - K_k\Phi_k.
\end{equation}

The noise-driven part then evolves as
\begin{align*}
    \zeta_{k+1} &= \Tilde{x}_{k+1|k} - \Gamma_{k+1}\Delta\theta \\
    &= (A - K_k C)\zeta_k + w_k - K_k v_k \\
     &+ \Big( (A - K_k C)\Gamma_k + \Xi_k - K_k\Phi_k - \Gamma_{k+1} \Big)\Delta\theta.
      \numberthis \label{eq:zeta_plus} \\
    &= (A - K_k C)\zeta_k + w_k - K_k v_k, \numberthis \label{eq:zeta_next}
\end{align*}
where the last term in \eqref{eq:zeta_plus} equals zero, due to \eqref{eq:Gamma_next}.
It then becomes clear, that by choosing the initial value as $\zeta_0 = \tilde{x}_0$, \eqref{eq:zeta_next} equals exactly the state error for the nominal system. 
The innovation sequence for the faulty case then becomes
\begin{align}
    \epsilon_k &=\epsilon_{h,k} + \big(C\Gamma_k + \Phi_k\big)\Delta\theta, \label{eq:additive_fault}
\end{align}
where $\epsilon_{h,k}$ denotes the innovation sequence of the same Kalman filter applied to the nominal system. Hence the effect of the fault on the innovation is additive. It is governed by the matrix gain
$C\Gamma_k + \Phi_k$, which is computed recursively through \eqref{eq:Gamma_next}. This gain depends on the fault profile matrices $\Psi_k$, $\Upsilon_k$ and $\Phi_k$, and therefore
on the past inputs, but not on the fault vector $\Delta\theta$.

\subsection{Hypothesis testing}\label{sec:GLR}

Due to the additive fault in \eqref{eq:faulty_system}, a fault-free and a faulty hypothesis for the \gls{glr} test are given as
\begin{align}
    \mathbb{H}_0: \epsilon_k &\sim  \mathcal{N}(0,\Sigma_{k|k-1}) \\ 
    \mathbb{H}_1: \epsilon_k &\sim \mathcal{N}\big((C\Gamma_k + \Phi_k)\Delta\theta, \Sigma_{k|k-1}\big),
\end{align}
where $\Sigma_{k|k-1}$ is the innovation covariance related to the nominal filter.
Since the fault vector $\Delta\theta$ is unknown under $\mathbb{H}_1$, it is replaced by its maximum likelihood estimator (MLE) based on the innovations up to time $k$
\begin{equation}\label{eq:mle}
    \Delta\widehat{\theta}_k = G_k^{-1}\, d_k,
\end{equation}
where
\begin{align}
    G_k &= \sum_{j=1}^{k} (C\Gamma_j + \Phi_j)^\top \Sigma_{j|j-1}^{-1} (C\Gamma_j + \Phi_j), \label{eq:G_k}\\
    d_k &= \sum_{j=1}^{k} (C\Gamma_j + \Phi_j)^\top \Sigma_{j|j-1}^{-1} \epsilon_j . \label{eq:d_k}
\end{align}
The \gls{glr} test statistics $l_k$ are defined as log-likelihood ratio between $\mathbb{H}_1$, with $\Delta\theta$ replaced by its MLE \eqref{eq:mle}, and $\mathbb{H}_0$, with $\Delta\theta = 0$
\begin{align*}
    l_k &\triangleq 2 \ln \frac{p\big(\epsilon_1,\dots,\epsilon_k \mid \Delta\theta = \Delta\widehat{\theta}_k\big)}{p\big(\epsilon_1,\dots,\epsilon_k \mid \Delta\theta = 0\big)}
    = d_k^\top G_k^{-1} d_k .
\end{align*}
The statistic $l_k$ is compared to a threshold $T$ for deciding between $\mathbb{H}_0$ and $\mathbb{H}_1$, $l_k \;\overset{\mathbb{H}_1}{\underset{\mathbb{H}_0}{\gtrless}}\; T$. It is known \cite[]{book} that $l_k$ follows a $\chi^2$ distribution, with noncentrality parameter $\lambda=0$ under $\mathbb{H}_0$ and noncentrality parameter $\lambda > 0$ under $\mathbb{H}_1$.

\section{Designing Optimal Inputs}\label{sec:OptimalInput}
For a fixed threshold $T$, a noncentral $\chi^2$ variable with fixed degrees of freedom has a detection probability that increases with its noncentrality parameter $\lambda$ \cite[]{book}.
From \eqref{eq:additive_fault}, \eqref{eq:G_k}, and \eqref{eq:d_k}, the noncentrality parameter can be written as $\lambda_k=\Delta\theta^\top G_k\Delta\theta$.

From \eqref{eq:G_k}, it follows that the noncentrality parameter accumulates over time and can be defined recursively as
\begin{align}
    \lambda_{k+1}
    &= \Delta\theta^\top G_{k+1}\Delta\theta\\
    &= \lambda_k+
    (L_{k+1}\Delta\theta)^\top
    \Sigma_{k+1|k}^{-1}L_{k+1}\Delta\theta,
    \label{eq:lambda_rec}
\end{align}
where $L_{k+1}=C\Gamma_{k+1}+\Phi_{k+1}$, and
\begin{equation}\label{eq:Gamma_deta_rec}
\begin{aligned}
    \Gamma_{k+1}\Delta\theta
    &= (A-K_kC)\Gamma_k\Delta\theta\\
    &\quad+(\Psi_k-K_k\Phi_k)\Delta\theta
    +\Upsilon_k\Delta\theta.
\end{aligned}
\end{equation}
Furthermore, the innovation sequence for the faulty case has the mean
\begin{equation}\label{eq:mean_explicit}
    \mu_k=(C\Gamma_k+\Phi_k)\Delta\theta
    =C\Gamma_k\Delta\theta+\Phi_k\Delta\theta.
\end{equation}
Note that the input affects the noncentrality directly through $\Upsilon_k$, and indirectly through the state estimate $\widehat{x}_{k|k-1}$ contained in $\Psi_k$ and $\Phi_k$.

The change vector $\Delta\theta$ is in principle unknown. We hereafter assume that the fault belongs to a finite set of candidate changes $\mathcal{F}=\big\{\Delta\theta^{(1)},\dots,\Delta\theta^{(n_f)}\big\}$. One typical choice is the set of single-element changes. Specifically, $\Delta\theta^{(f)}=-\delta_f e_f$, $f=1,\dots,p$, where $e_f\in\mathbb{R}^p$ is the $f$-th unit vector and $\delta_f>0$. 

Since the true fault location is unknown, the input is designed to maximize the worst-case expected noncentrality over $\mathcal{F}$ and a finite horizon $H$. Denoting the input sequence for a horizon $H$ by $U_k=[u_k^\top\ \cdots\ u_{k+H-1}^\top]^\top$, the optimization problem we are interested in becomes
\begin{equation}\label{eq:maxmin_equiv}
\begin{aligned}
\max_{U_k}\ \min_{f\in\mathcal{F}}\quad&
\mathbb{E}\big[
\lambda_{k+H}(\Delta\theta^{(f)})
-\lambda_k(\Delta\theta^{(f)})
\mid\mathcal{I}_k\big]\\
\text{s.t.}\quad&
\widehat{x}_{j+1|j}
=A\widehat{x}_{j|j-1}+B\bar u_j+K_j\epsilon_j,\\
&\Gamma_{j+1}
=(A-K_jC)\Gamma_j+\Psi_j+\Upsilon_j-K_j\Phi_j,\\
&\|u_j\|_2^2\le\bar P,
\qquad j=k,\dots,k+H-1,
\end{aligned}
\end{equation}
where $\mathcal{I}_k$ denotes the information available at time $k$, consisting of the previous measurements $y_0,\dots,y_{k-1}$, applied inputs $u_0,\dots,u_{k-1}$, and measured operational inputs $z_0,\dots,z_k$.

Note that the future innovations $\epsilon_j$ are random, and therefore the predicted state estimates, fault sensitivities, and noncentrality parameters are also random. 
In this paper, the expected noncentrality parameter is used when designing the input to obtain a tractable  deterministic problem. More specifically, for each candidate fault $\Delta\theta^{(f)}$, let $\mu_j^{(f)}=(C\Gamma_j+\Phi_j)\Delta\theta^{(f)}$ denote the fault-induced innovation mean. Defining its conditional mean and covariance as
\begin{equation}
    \bar\mu_{j|k}^{(f)}
    =\mathbb{E}[\mu_j^{(f)}\mid\mathcal{I}_k], \quad
    V_{j|k}^{(f)}
    =\operatorname{Cov}(\mu_j^{(f)}\mid\mathcal{I}_k),
\end{equation}
the expected increase in the noncentrality parameter can then be expressed as
\begin{multline}\label{eq:expected_lambda}
\mathbb{E}\big[
\lambda_{k+H}(\Delta\theta^{(f)})
-\lambda_k(\Delta\theta^{(f)})
\mid\mathcal{I}_k\big]\\
=\sum_{j=k+1}^{k+H}
\Big[
(\bar\mu_{j|k}^{(f)})^\top
\Sigma_{j|j-1}^{-1}\bar\mu_{j|k}^{(f)}
+\operatorname{tr}\big(
\Sigma_{j|j-1}^{-1}V_{j|k}^{(f)}
\big)
\Big].
\end{multline}
Since the covariance term $\operatorname{tr}\big(
\Sigma_{j|j-1}^{-1}V_{j|k}^{(f)}
\big)$ is nonnegative for any $j\in\{k+1,\ldots,k+H\}$, it follows that
\begin{multline}\label{eq:lambda_lower_bound}
\mathbb{E}\big[\lambda_{k+H}(\Delta\theta^{(f)})-\lambda_k(\Delta\theta^{(f)}) \mid\mathcal{I}_k\big]\\\geq
\sum_{j=k+1}^{k+H}(\bar\mu_{j|k}^{(f)})^\top
\Sigma_{j|j-1}^{-1}\bar\mu_{j|k}^{(f)}.
\end{multline}

For simplicity and numerical efficiency the lower bound $\sum_{j=k+1}^{k+H}(\bar\mu_{j|k}^{(f)})^\top
\Sigma_{j|j-1}^{-1}\bar\mu_{j|k}^{(f)}$ is maximized instead of the expected noncentrality parameter.

Let $\bar x_{j|k}$ and $\bar\Gamma_{j|k}$ denote the predicted state and fault sensitivity, respectively, initialized at $\bar x_{k|k}=\widehat{x}_{k|k-1}$ and $\bar\Gamma_{k|k}=\Gamma_k$. The conditional means $\bar\mu_{j|k}^{(f)}$ are then obtained by propagating these predictions using $\mathbb{E}[\epsilon_j\mid\mathcal{I}_k]=0$ under the nominal model for $j\geq k$. The resulting deterministic optimization problem is
\begin{equation}\label{eq:det_horizon}
\begin{aligned}
\max_{U_k}\ \min_{f\in\mathcal{F}}\quad&
\sum_{j=k+1}^{k+H}
(\bar\mu_{j|k}^{(f)})^\top
\Sigma_{j|j-1}^{-1}\bar\mu_{j|k}^{(f)}\\
\text{s.t.}\quad&
\bar x_{j+1|k}
=A\bar x_{j|k}+B(u_j+\check z_{j|k}),\\
&\bar\Gamma_{j+1|k}
=(A-K_jC)\bar\Gamma_{j|k}\\
&\qquad+\Psi_j(\bar x_{j|k})
-K_j\Phi_j(\bar x_{j|k})\\
&\qquad+\Upsilon_j(u_j+\check z_{j|k}),\\
&\|u_j\|_2^2\le\bar P,
\qquad j=k,\dots,k+H-1,\\
&\bar\mu_{j|k}^{(f)}
=(C\bar\Gamma_{j|k}+\Phi_j(\bar x_{j|k}))
\Delta\theta^{(f)},\\
&\hspace{3cm}j=k+1,\dots,k+H,
\end{aligned}
\end{equation}
where $\check z_{j|k}$ is the available or the predicted uncontrolled input.
Maximizing the detectability of a specific fault scenario is then a special case $n_f=1$ with $\Delta\theta^{(1)}=\Delta\theta_0$.

Subsequently, we assume that the experiment or the system  of interest is fixed to run for a specific known horizon $N\geq H>0$. Given $N$, we calculate the inputs using three different approaches:  an offline or an open-loop approach where the input sequence is computed once and applied for a fixed horizon $H=N$ in \eqref{eq:det_horizon}, a \gls{rh} approach where \eqref{eq:det_horizon} is solved repeatedly for $H<N$ every time we obtain a new measurement, and a \gls{sh} approach \cite[]{rawlings2020model}, in which $H$ is adjusted the closer we get to $N$.

\subsection{Offline (Open-Loop) Input Design}\label{sec:offline}

In the offline case, \eqref{eq:det_horizon} is solved at $k=0$ with $H=N$, without using measurements collected during the experiment. We write $U=U_0$, $\bar x_j=\bar x_{j|0}$, and $\bar\Gamma_j=\bar\Gamma_{j|0}$ for $j=0,\dots,N$, and assume $\bar x_0=0$ and $\bar\Gamma_0=0$. The latter means that the fault has not yet affected the state error at time zero.

As for the measured input $z_j$, it is assumed to be constant in the offline optimization, i.e., $\hat z_{j|0}=\hat z$ for all $j=0,\dots,N-1$. We denote the corresponding vector by $Z=[\hat z^\top\ \cdots\ \hat z^\top]^\top\in\mathbb{R}^{Nl}$.

Following \cite[]{dohler2017,GRES2023}, the linear dependence of $\Psi_j(\bar x_j)$ and $\Phi_j(\bar x_j)$ on $\bar x_j$, and of $\Upsilon_j(u_j+\hat z)$ on $u_j+\hat z$, allows the state and input to be factored from the fault signatures in \eqref{eq:det_horizon}.
Thus, stacking the predicted innovation means in \eqref{eq:det_horizon} for $j\in\{1,\dots,N\}$ gives
\begin{equation}\label{eq:toeplitz}
    \begin{bmatrix}
        \bar\mu_{1|0}^{(f)}\\
        \bar\mu_{2|0}^{(f)}\\
        \vdots\\
        \bar\mu_{N|0}^{(f)}
    \end{bmatrix}
    =\mathcal{T}(\Delta\theta^{(f)})(U+Z),
\end{equation}
where $\mathcal{T}(\Delta\theta^{(f)})\in\mathbb{R}^{Nr\times Nl}$ is a block lower-triangular matrix of the form
\begin{equation}\label{eq:low_triang}
    \mathcal{T}(\Delta\theta^{(f)})=
    \begin{bmatrix}
        \mathcal{T}_{1,0} & 0 & \cdots & 0\\
        \mathcal{T}_{2,0} & \mathcal{T}_{2,1} & \ddots & \vdots\\
        \vdots & \vdots & \ddots & 0\\
        \mathcal{T}_{N,0} & \mathcal{T}_{N,1} & \cdots & \mathcal{T}_{N,N-1}
    \end{bmatrix}.
\end{equation}

The blocks in \eqref{eq:low_triang} follow from the prediction dynamics in \eqref{eq:det_horizon}. The matrix $\mathcal{T}(\Delta\theta^{(f)})$ is block lower triangular, since $u_i$ first appears in the predicted innovation mean at time $i+1$. If the Kalman gain $K_k$ is constant (i.e. if the stationary Kalman gain is used) and the fault sensitivity mappings are time-invariant, then $\mathcal{T}(\Delta\theta^{(f)})$ becomes block Toeplitz, which can be used to considerably improve the numerical computation time.

As the lower bound in \eqref{eq:lambda_lower_bound} depends on the innovation covariances, these are collected in a matrix
\begin{equation*}
    \mathcal{W}=\mathrm{blockdiag}
    \big(\Sigma_{1|0}^{-1},\dots,\Sigma_{N|N-1}^{-1}\big).
\end{equation*}
The lower bound in \eqref{eq:lambda_lower_bound} can then be written as
\begin{equation}\label{eq:Elambda}
    \sum_{j=1}^{N}
    (\bar\mu_{j|0}^{(f)})^\top
    \Sigma_{j|j-1}^{-1}\bar\mu_{j|0}^{(f)}
    =(U+Z)^\top\mathcal{Q}(\Delta\theta^{(f)})(U+Z),
\end{equation}
where $\mathcal{Q}(\Delta\theta^{(f)})=\mathcal{T}(\Delta\theta^{(f)})^\top\mathcal{W}\mathcal{T}(\Delta\theta^{(f)})$. The offline input is then the solution of
\begin{equation}\label{eq:offline_problem}
\begin{aligned}
    \max_U\ \min_{f=1,\dots,n_f}\quad&
    (U+Z)^\top\mathcal{Q}(\Delta\theta^{(f)})(U+Z)\\
    \text{s.t.}\quad&
    \|u_j\|_2^2\le\bar P,\qquad j=0,\dots,N-1.
\end{aligned}
\end{equation}

Since $\mathcal{Q}(\Delta\theta^{(f)})$ is positive semidefinite, each candidate objective is convex in $U$. For a known fault, i.e., $n_f=1$, the constraints define a compact convex set, so an optimal solution can be chosen at an extreme point, i.e., $\|u_j\|_2^2=\bar P$ for all $j$. For multiple candidate faults, this is not necessarily the case. The problem is solved by a \gls{mm} algorithm \cite[]{alma9920637831505762}, where each candidate objective is linearised at the current iteration. By convexity, this gives a lower bound that is maximised, so that each iteration does not decrease the objective.

\subsection{Online Receding-Horizon Design}\label{sec:online}

While the offline design optimizes the entire sequence $U$ before the experiment, the online policy adapts to the measured data. Specifically, at each time step $k$, the current nominal Kalman filter state estimate $\widehat{x}_{k|k-1}$, the accumulated fault sensitivity $\Gamma_k$, and the measured input $z_k$ are known. These are used in \eqref{eq:det_horizon} to predict the model's behavior over the next $H$ steps.

Utilizing the same approach as in Section \ref{sec:offline}, stacking the predicted innovation means in \eqref{eq:det_horizon} over the horizon $H$ gives
\begin{equation}\label{eq:online_pred_stack}
    \begin{bmatrix}
        \bar\mu_{k+1|k}^{(f)}\\
        \vdots\\
        \bar\mu_{k+H|k}^{(f)}
    \end{bmatrix}
    =a_k(\Delta\theta^{(f)})
    +\mathcal{T}_H(\Delta\theta^{(f)})(U_k+Z_k),
\end{equation}
where $Z_k=[\hat z_{k|k}^\top\ \cdots\ \hat z_{k+H-1|k}^\top]^\top\in\mathbb{R}^{Hl}$ contains the available or predicted operational inputs, with $\hat z_{k|k}=z_k$. The term $a_k(\Delta\theta^{(f)})$ represents the stacked prediction obtained from \eqref{eq:det_horizon} with the future designed and operational inputs set to zero, and therefore, it depends on $\widehat{x}_{k|k-1}$ and $\Gamma_k$ only.

The matrix $\mathcal{T}_H(\Delta\theta^{(f)})\in\mathbb{R}^{Hr\times Hl}$ is block lower triangular, with a structure similar to \eqref{eq:low_triang}, over the horizon $H$ at time $k$.
The online input is then computed as the solution of the following optimization problem at each step $k$
\begin{table*}[h!]
\centering
\caption{Monte Carlo evaluation (100 runs) of \gls{glr} detection metrics for a stationary load.}
\label{tab:mc_nodrift}
\small
\setlength{\tabcolsep}{4pt}
\setlength{\aboverulesep}{1.5pt}   
\setlength{\belowrulesep}{1.5pt}
\resizebox{\textwidth}{!}{%
\begin{tabular}{@{}l ccc ccc ccc ccc@{}}
\toprule
\multirow{4}{*}{\textbf{Policy}}
 & \multicolumn{6}{c}{\textbf{$\mathbf{0.5\%}$ fault}} & \multicolumn{6}{c}{\textbf{$\mathbf{5\%}$ fault}} \\
\cmidrule(lr){2-7}\cmidrule(lr){8-13}
 & \multicolumn{3}{c}{$k_{\mathrm{f}}=150$} & \multicolumn{3}{c}{$k_{\mathrm{f}}=400$}
 & \multicolumn{3}{c}{$k_{\mathrm{f}}=150$} & \multicolumn{3}{c}{$k_{\mathrm{f}}=400$} \\
\cmidrule(lr){2-4}\cmidrule(lr){5-7}\cmidrule(lr){8-10}\cmidrule(lr){11-13}
 & Delay & $P_D$ & $\bar\lambda_N$ & Delay & $P_D$ & $\bar\lambda_N$ & Delay & $P_D$ & $\bar\lambda_N$ & Delay & $P_D$ & $\bar\lambda_N$ \\
\midrule
Passive  & $320.1 \pm 137.2$ & $62\%$ & $27.4$ & $273.2 \pm 59.1$ & $36\%$ & $27.3$ & $30.4 \pm 16.6$ & $100\%$ & $3429$ & $36.9 \pm 22.9$ & $100\%$ & $3797$ \\
Offline                & $164.5 \pm 80.2$ & $\mathbf{100\%}$ & $82.4$ & $202.5 \pm 74.8$ & $95\%$ & $82.1$ & $18.2 \pm 6.1$ & $100\%$ & $9490$ & $\mathbf{17.0 \pm 4.6}$ & $100\%$ & $10926$ \\
Greedy ($H=1$)      & $316.7 \pm 140.2$ & $90\%$ & $46.1$ & $273.0 \pm 73.8$ & $54\%$ & $46.0$ & $19.8 \pm 6.5$ & $100\%$ & $6095$ & $26.8 \pm 9.8$ & $100\%$ & $6343$ \\
RH ($H=10$)         & $249.9 \pm 126.3$ & $98\%$ & $60.5$ & $239.6 \pm 95.0$ & $68\%$ & $60.3$ & $\mathbf{16.2 \pm 5.1}$ & $100\%$ & $8873$ & $21.8 \pm 7.1$ & $100\%$ & $9758$ \\
SH ($H=N-k$) & $\mathbf{158.4 \pm 69.2}$ & $\mathbf{100\%}$ & $\mathbf{106.1}$ & $\mathbf{176.3 \pm 69.2}$ & $\mathbf{99\%}$ & $\mathbf{105.5}$ & $17.6 \pm 5.4$ & $100\%$ & $\mathbf{12834}$ & $\mathbf{17.0 \pm 4.8}$ & $100\%$ & $\mathbf{14870}$ \\
\bottomrule
\end{tabular}}
\end{table*}
\begin{equation}\label{eq:online_problem}
\begin{aligned}
    \max_{U_k}\ \min_{f=1,\dots,n_f}\quad&
    \big[a_k(\Delta\theta^{(f)})
    +\mathcal{T}_H(\Delta\theta^{(f)})(U_k+Z_k)\big]^\top\\
    &\mathcal{W}_H
    \big[a_k(\Delta\theta^{(f)})
    +\mathcal{T}_H(\Delta\theta^{(f)})(U_k+Z_k)\big]\\
    \text{s.t.}\quad&
    \|u_{k+i}\|_2^2\le\bar P,\qquad i=0,\dots,H-1,
\end{aligned}
\end{equation}
where
\begin{equation*}
    \mathcal{W}_H=\mathrm{blockdiag}
    \big(\Sigma_{k+1|k}^{-1},\dots,
    \Sigma_{k+H|k+H-1}^{-1}\big), 
\end{equation*}
and the applied input $u_k$ is the first input of $U_k$.
Problem \eqref{eq:online_problem} has the same quadratic structure for each candidate fault as \eqref{eq:offline_problem}, and is solved by the same \gls{mm} algorithm.

\section{Numerical Results}\label{sec:results}

In this section, the proposed scheme is applied in a numerical experiment on a 6 DOF mechanical chain-like system that for any consistent set of units is modeled with spring stiffness $k_1 = k_3 = k_5 = 100$ and $k_2 = k_4 = k_6 = 200$, mass $m_i = 1/20$ and a proportional damping matrix such that all modes have a damping ratio of 2\%. The system is excited by periodic inputs acting at all DOFs. 
The accelerations of all six masses are measured at a sampling frequency of $50$~Hz for 16 seconds.
The force on
each mass is the sum of the designed input, the measured operational load and unmeasured white process noise with standard deviation $0.1$~N. White measurement noise with standard deviation $0.01$ $\tfrac{m}{s^2}$ is added to each acceleration measurement.

We consider and compare five control input design strategies: a passive approach, where no input is added, the offline approach described in Section \ref{sec:offline}, and the online \gls{rh} approach, described in Section \ref{sec:online}, with horizon $H=1,10,$ and $N-k$, i.e. greedy, \gls{rh} and \gls{sh}, respectively.
One threshold $T$ is calibrated per strategy, over 200 fault-free Monte Carlo iterations, to ensure that the false alarm rate for the \gls{glr} test is at $5\%$. $T$ is then utilized to normalize the \gls{glr} test.

For each input design strategy, two cases are considered.
In case 1, the operational load is a sum of three sinusoids with constant frequencies, $0.3, 1.9, 9.0$Hz, amplitudes $0.5,0.25,0.25$N and phases which are uniform on $[0,2\pi)$.
In case 2, the operational load is once again modeled as described in case 1 with the three sinusoids, but with drifting frequencies, which increase linearly with $0.5$Hz/s. The drift is introduced to simulate realistic operating environments where a system's external excitation naturally varies over time.

Each strategy is evaluated using a Monte Carlo simulation, with 100 iterations, with the same noise realizations, by comparing the detection probability ($P_D$), the mean detection delay, and the mean noncentrality parameter at time step $N$ ($\overline{\lambda}_N$). Since the mean delay is computed only when the fault is detected, delays are compared only between strategies with similar detection probability $P_D$.
To ensure safe operation of the system, a constraint $\|u_k\|^2_2\le 0.05$ is imposed on the power of the input.

\subsection{Results Known Fault}\label{sec:known_fault_results}
First consider the special case where the specific fault scenario is known, given by a simultaneous loss of stiffness in springs 3-5 of either $0.5\%$ or $5\%$, starting at $k_{\mathrm f}=150$ or $k_{\mathrm f}=400$.

\begin{figure}[h!]
  \centering
  \resizebox{\columnwidth}{!}{\includegraphics[page=6, width=1\textwidth]{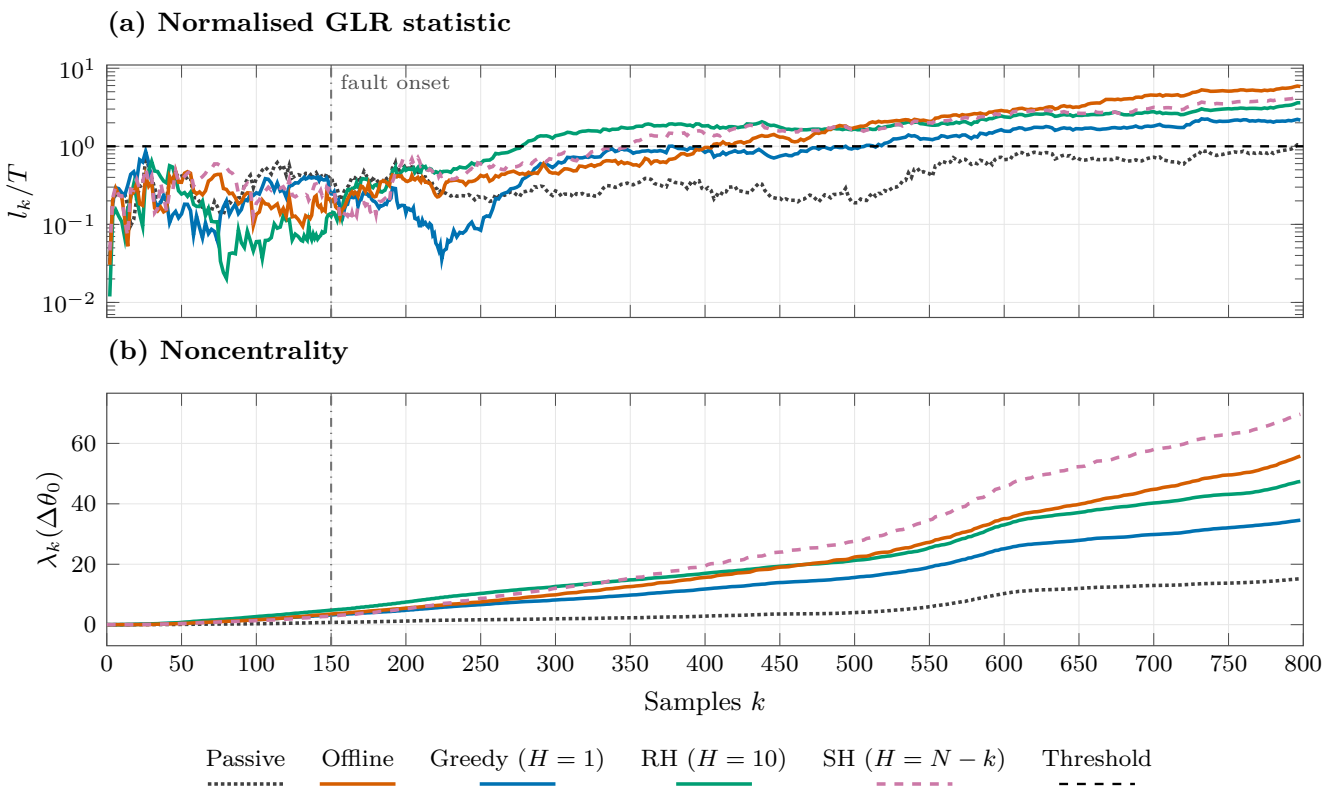}}
  \caption{Known fault case 1. One realization for a 0.5\% stiffness loss at $k_{\mathrm f}=150$, marked with a vertical line. Normalised \gls{glr} (a) and noncentrality (b).
  }
  \label{fig:glr}
\end{figure}

For case 1, Fig. \ref{fig:glr} shows that the designed inputs make the noncentrality grow considerably faster than passive excitation, so the \gls{glr} statistic crosses the threshold much earlier.
This is further confirmed when looking at Table \ref{tab:mc_nodrift} showing the Monte Carlo results for case 1. The passive approach is outperformed in all metrics, with the \gls{sh} strategy reaching a detection probability of $\approx100\%$ for all scenarios, while consistently having the lowest delay. Offline and \gls{sh} perform similarly, with \gls{sh} better only for the small, late fault. Greedy and \gls{rh} are competitive for the $5\%$ fault but detect the $0.5\%$ fault less often, since a small fault only becomes visible once information has accumulated over many samples, which favours maximizing the noncentrality over the remaining experiment.
For case 2, only the $0.5\%$ fault is reported since this is where the results differ the most. It is seen in Table~\ref{tab:mc_results_drift}, that \gls{sh} outperforms offline, reducing the delay from $334$ to $195$ samples at $k_{\mathrm f}=150$ and raising $P_D$ from $85\%$ to $98\%$ at $k_{\mathrm f}=400$.
The offline input, that only has knowledge of the first 100 samples, falls out of phase with the drifting load, whereas the online strategies measure the load and re-plan at every step.
\begin{table}[h!]
\centering
\caption{Monte Carlo evaluation (100 runs) with 0.5\,Hz/s load frequency drift, 0.5\% fault.}
\label{tab:mc_results_drift}
\small
\setlength{\tabcolsep}{4pt}
\setlength{\aboverulesep}{1.5pt}   
\setlength{\belowrulesep}{1.5pt}
\begin{tabular}{@{}l ccc ccc@{}}
\toprule
 & \multicolumn{3}{c}{$k_{\mathrm{f}}=150$} & \multicolumn{3}{c}{$k_{\mathrm{f}}=400$} \\
\cmidrule(lr){2-4}\cmidrule(lr){5-7}
Policy & Delay & $P_D$ & $\bar\lambda_N$ & Delay & $P_D$ & $\bar\lambda_N$ \\
\midrule
Passive & $421 \pm 169$ & $44\%$  & $13.5$ & $278 \pm 101$ & $33\%$ & $13.5$ \\
Offline & $334 \pm 126$ & $\mathbf{100\%}$ & $46.5$ & $239 \pm 81$  & $85\%$ & $46.5$ \\
Greedy  & $333 \pm 150$ & $96\%$  & $32.4$ & $261 \pm 88$  & $50\%$ & $32.4$ \\
RH      & $312 \pm 143$ & $99\%$  & $44.1$ & $265 \pm 91$  & $63\%$ & $44.2$ \\
SH      & $\mathbf{195 \pm 98}$ & $\mathbf{100\%}$ & $\mathbf{69.0}$ & $\mathbf{218 \pm 74}$ & $\mathbf{98\%}$ & $\mathbf{69.1}$ \\
\bottomrule
\end{tabular}
\end{table}

\begin{table*}[h!]
\centering
\caption{Monte Carlo evaluation (120 runs) for an unknown fault location with 0.5 Hz/s load frequency drift.}
\label{tab:mc_unknown}
\small
\setlength{\tabcolsep}{4pt}
\setlength{\aboverulesep}{2pt}   
\setlength{\belowrulesep}{2pt}
\resizebox{\textwidth}{!}{%
\begin{tabular}{@{}l ccc ccc ccc ccc@{}}
\toprule
\multirow{4}{*}{\textbf{Policy}}
 & \multicolumn{6}{c}{\textbf{$\mathbf{0.5\%}$ fault}} & \multicolumn{6}{c}{\textbf{$\mathbf{5\%}$ fault}} \\
\cmidrule(lr){2-7}\cmidrule(lr){8-13}
 & \multicolumn{3}{c}{$k_{\mathrm{f}}=150$} & \multicolumn{3}{c}{$k_{\mathrm{f}}=400$}
 & \multicolumn{3}{c}{$k_{\mathrm{f}}=150$} & \multicolumn{3}{c}{$k_{\mathrm{f}}=400$} \\
\cmidrule(lr){2-4}\cmidrule(lr){5-7}\cmidrule(lr){8-10}\cmidrule(lr){11-13}
 & Delay & $P_D$ & $\min_f\bar\lambda_N$ & Delay & $P_D$ & $\min_f\bar\lambda_N$ & Delay & $P_D$ & $\min_f\bar\lambda_N$ & Delay & $P_D$ & $\min_f\bar\lambda_N$ \\
\midrule
Passive & $338.1 \pm 181.7$ & $15\%$ & $2.6$ & $188.0 \pm 113.5$ & $10\%$ & $2.6$ & $97.0 \pm 107.6$ & $100\%$ & $293$ & $73.2 \pm 61.5$ & $99\%$ & $288$ \\
Offline & $414.9 \pm 163.2$ & $45\%$ & $9.0$ & $236.6 \pm 93.6$ & $24\%$ & $9.1$ & $40.1 \pm 29.7$ & $100\%$ & $929$ & $26.1 \pm 15.7$ & $100\%$ & $992$ \\
Greedy ($H=1$) & $366.1 \pm 198.3$ & $28\%$ & $4.2$ & $230.0 \pm 122.2$ & $12\%$ & $4.1$ & $59.1 \pm 54.0$ & $100\%$ & $461$ & $57.4 \pm 45.3$ & $100\%$ & $453$ \\
RH ($H=10$) & $416.7 \pm 178.3$ & $28\%$ & $6.1$ & $222.6 \pm 125.0$ & $13\%$ & $6.1$ & $38.2 \pm 22.8$ & $100\%$ & $693$ & $56.1 \pm 40.8$ & $100\%$ & $659$ \\
SH ($H=N-k$) & $\mathbf{345.5 \pm 169.1}$ & $\mathbf{68\%}$ & $\mathbf{16.0}$ & $\mathbf{226.3 \pm 100.4}$ & $\mathbf{37\%}$ & $\mathbf{16.0}$ & $\mathbf{23.9 \pm 19.3}$ & $100\%$ & $\mathbf{1727}$ & $\mathbf{23.3 \pm 15.2}$ & $100\%$ & $\mathbf{1668}$ \\
\midrule
Oracle, offline & $391.2 \pm 154.1$ & $77\%$ & $4.3$ & $266.7 \pm 80.0$ & $42\%$ & $4.3$ & $21.7 \pm 13.6$ & $100\%$ & $436$ & $24.6 \pm 14.7$ & $100\%$ & $445$ \\
Oracle, SH ($H=N-k$) & $349.1 \pm 167.4$ & $88\%$ & $5.9$ & $255.8 \pm 82.5$ & $56\%$ & $5.9$ & $16.5 \pm 9.2$ & $100\%$ & $610$ & $22.2 \pm 12.9$ & $100\%$ & $614$ \\
\bottomrule
\end{tabular}}
\end{table*}

\subsection{Results Unknown Fault}
For the unknown fault scenario, in each Monte Carlo iteration, one spring is damaged, and the damaged spring is cycled through all six springs, so each spring is the true fault in $20$ of the $120$ runs. No strategy knows which spring is damaged.
To measure the cost of not knowing the fault location, we include two references that know the damaged spring, an offline oracle and a \gls{sh} oracle. 
Note that in Section \ref{sec:known_fault_results}, the fault is $0.5\%$ on each of the three springs, while here it is $0.5\%$ on a single spring. Thus, the accumulated fault across the system is lower, which explains the lower $P_D$ of the oracles reported in Table \ref{tab:mc_unknown}.
Since the objective is to maximize the worst-case over $\mathcal F$, Table~\ref{tab:mc_unknown} reports the minimum noncentrality 
$\min_f\lambda_N(\Delta\theta^{(f)})$.
As shown in Table \ref{tab:mc_unknown} when the fault location is unknown, the passive approach fails to provide meaningful detection for the 0.5\% fault. Also the offline, greedy and \gls{rh} perform poorly with a detection probability below $45\%$.
It is again seen that the \gls{sh} outperforms the other inputs for all scenarios. 
In general, when comparing the oracle offline and oracle \gls{sh} to their non-oracle counterpart, it is seen that they both suffer in either the detection probability or in the delay of detection. However it is seen that the minimum noncentrality is higher, thus if a fault happened elsewhere, the oracle would perform worse.

\section{Conclusion}

This paper proposed an active input design for statistical detection of small parametric faults in stochastic \gls{lti} systems. The input maximizes the noncentrality parameter of a \gls{glr} test on the innovations of the nominal Kalman filter. The performance of offline and online designs was evaluated on a six-degree-of-freedom mass-spring chain under constrained input power. All designed inputs reduce detection delays and increases detection probabilities compared to passive excitation. The online \gls{sh} policy performed well across various scenarios, outperforming the offline design under non-stationary operating conditions, by adapting along the way. Furthermore, the \gls{sh} approach maintained superior worst-case detectability even when the exact location of the fault was unknown.

\bibliography{ifacconf}

@article{GRES2023,
title = {Parameter-state estimation for mechanical systems with small model errors},
journal = {IFAC-PapersOnLine},
volume = {56},
number = {2},
pages = {10264-10269},
year = {2023},
note = {22nd IFAC World Congress},
issn = {2405-8963},
author = {Szymon Greś and Konstantinos Tatsis and Vasilis Dertimanis and Eleni Chatzi},
}

@article{zhang2014,
title = {Statistical detection and isolation of additive faults in linear time-varying systems},
journal = {Automatica},
volume = {50},
number = {10},
pages = {2527-2538},
year = {2014},
issn = {0005-1098},
author = {Qinghua Zhang and Michèle Basseville}
}

@article{dohler2017,
title = {Change detection and isolation in mechanical system parameters based on perturbation analysis},
journal = {IFAC-PapersOnLine},
volume = {50},
number = {1},
pages = {419-424},
year = {2017},
note = {20th IFAC World Congress},
issn = {2405-8963},
author = {Michael Döhler and Qinghua Zhang and Laurent Mevel},
}

@article{niemann2006setup,
  title={A setup for active fault diagnosis},
  author = {Hans Henrik Niemann},
  journal={IEEE Transactions on Automatic Control},
  volume={51},
  number={9},
  pages={1572--1578},
  year={2006},
  publisher={IEEE}
}

@book{book,
author = {Basseville, Michèle and Nikiforov, Igor},
year = {1993},
month = {04},
pages = {},
title = {Detection of Abrupt Change Theory and Application},
volume = {15},
isbn = {0-13-126780-9},
}

@article{gao2015survey,
  title={A survey of fault diagnosis and fault-tolerant techniques—Part I: Fault diagnosis with model-based and signal-based approaches},
  author={Gao, Zhiwei and Cecati, Carlo and Ding, Steven X},
  journal={IEEE transactions on industrial electronics},
  volume={62},
  number={6},
  pages={3757--3767},
  year={2015},
  publisher={IEEE}
}

@ARTICLE{6179529,
  author={Helm, Siegfried and Kozek, Martin and Jakubek, Stefan},
  journal={IEEE Transactions on Industrial Electronics}, 
  title={Combustion Torque Estimation and Misfire Detection for Calibration of Combustion Engines by Parametric Kalman Filtering}, 
  year={2012},
  volume={59},
  number={11},
  pages={4326-4337}
}

@ARTICLE{6135786,
  author={Lall, Pradeep and Lowe, Ryan and Goebel, Kai},
  journal={IEEE Transactions on Industrial Electronics}, 
  title={Prognostics Health Management of Electronic Systems Under Mechanical Shock and Vibration Using Kalman Filter Models and Metrics}, 
  year={2012},
  volume={59},
  number={11},
  pages={4301-4314}
}

@ARTICLE{6428675,
  author={Foo, Gilbert Hock Beng and Zhang, Xinan and Vilathgamuwa, D. M.},
  journal={IEEE Transactions on Industrial Electronics}, 
  title={A Sensor Fault Detection and Isolation Method in Interior Permanent-Magnet Synchronous Motor Drives Based on an Extended Kalman Filter}, 
  year={2013},
  volume={60},
  number={8},
  pages={3485-3495}
}

@article{SCOTT20141580,
title = {Input design for guaranteed fault diagnosis using zonotopes},
journal = {Automatica},
volume = {50},
number = {6},
pages = {1580-1589},
year = {2014},
issn = {0005-1098},
author = {Joseph K. Scott and Rolf Findeisen and Richard D. Braatz and Davide M. Raimondo},
}

@article{ESNAASHARI20111887,
title = {Auxiliary signal design for robust active fault detection of linear discrete-time systems},
journal = {Automatica},
volume = {47},
number = {9},
pages = {1887-1895},
year = {2011},
issn = {0005-1098},
author = {Alireza {Esna Ashari} and Ramine Nikoukhah and Stephen L. Campbell},
}

@article{raimondo2016closedloop,
title = {Closed-loop input design for guaranteed fault diagnosis using set-valued observers},
journal = {Automatica},
volume = {74},
pages = {107-117},
year = {2016},
issn = {0005-1098},
author = {Davide M. Raimondo and Giuseppe {Roberto Marseglia} and Richard D. Braatz and Joseph K. Scott},
}

@article{xu2021observer,
title = {Observer-based asymptotic active fault diagnosis: A two-layer optimization framework},
journal = {Automatica},
volume = {128},
pages = {109558},
year = {2021},
issn = {0005-1098},
author = {Feng Xu},
}

@INPROCEEDINGS{blackmore2006finite,
  author={Blackmore, L. and Williams, B.},
  booktitle={Proceedings of the 44th IEEE Conference on Decision and Control}, 
  title={Finite Horizon Control Design for Optimal Model Discrimination}, 
  year={2005},
  volume={},
  number={},
  pages={3795-3802}}

@INPROCEEDINGS{kim2013,
  author={Kim, Kwang-Ki K. and Raimondo, Davide M. and Braatz, Richard D.},
  booktitle={2013 European Control Conference (ECC)}, 
  title={Optimum input design for fault detection and diagnosis: Model-based prediction and statistical distance measures}, 
  year={2013},
  volume={},
  number={},
  pages={1940-1945}}

@article{rawlings2020model,
  title={Model predictive control: theory, computation, and design},
  author={Rawlings, James B and Mayne, David Q and Diehl, Moritz M},
  journal={Nob Hill Publishing, 2nd ed.},
  year={2020}
}

@book{alma9920637831505762,
author = {Ortega, J M.},
isbn = {0125285507},
title = {Iterative Solution of Nonlinear Equations in Several Variables.},
year = {1970},
}

\appendix

\end{document}